\documentclass{article}
\usepackage{ijcai25}

\usepackage{times}
\usepackage{soul}
\usepackage{url}
\usepackage[hidelinks]{hyperref}
\usepackage[utf8]{inputenc}
\usepackage[small]{caption}
\usepackage{graphicx}
\usepackage{amsmath}
\usepackage{amsthm}
\usepackage{amsfonts}
\usepackage{booktabs}
\usepackage{algorithm}
\usepackage{algpseudocode}
\usepackage[switch]{lineno}
\usepackage{xspace}
\usepackage{siunitx}
\usepackage{subcaption}
\usepackage{lipsum}

\usepackage{xcolor}
\usepackage{amssymb}
\usepackage{tikz}
\usetikzlibrary{patterns}
\usepackage{adjustbox}
\usepackage{multirow}
\usepackage{makecell}

\usepackage{pifont}%
\newcommand{\cmark}{\ding{51}}%
\newcommand{\xmark}{\ding{55}}%

\newcommand{\instrrst}[1]{{\scriptsize #1}}

\newcommand{\ie}{\textit{i.e.}~}
\newcommand{\eg}{\textit{e.g.}~}
\newcommand{\meteor}{\textsc{Meteor}\xspace}
\newcommand{\meteors}{\textsc{Meteor}s\xspace}
\newcommand{\figaro}{\textsc{Figaro}\xspace}
\newcommand{\token}[1]{{\tt#1}}
\makeatletter
\newcommand\footnoteref[1]{\protected@xdef\@thefnmark{\ref{#1}}\@footnotemark}
\makeatother

\makeatletter
\newcommand{\ssymbol}[1]{^{\@fnsymbol{#1}}}
\makeatother

\title{METEOR: Melody-aware Texture-controllable\\Symbolic Music Re-Orchestration via Transformer VAE}

\author{
    Dinh-Viet-Toan Le$^1$\And 
    Yi-Hsuan Yang$^2$\\
    \affiliations
    $^1$Univ. Lille, CNRS, Inria, Centrale Lille, UMR 9189 CRIStAL, F-59000 Lille, France\\
    $^2$National Taiwan University, Taiwan\\
    \emails
    dinhviettoan.le@univ-lille.fr,
    yhyangtw@ntu.edu.tw
}

\begin{document}

\maketitle

\begin{abstract}
Re-orchestration is the process of adapting a music piece for a different set of instruments. 
By altering the original instrumentation, the orchestrator often modifies the musical texture while preserving a recognizable melodic line and ensures that each part is playable within the technical and expressive capabilities of the chosen instruments.
In this work, we propose \meteor, a model for generating Melody-aware Texture-controllable re-Orchestration with a Transformer-based variational auto-encoder (VAE).
This model performs symbolic instrumental and textural music style transfers with a focus on melodic fidelity and controllability. 
We allow bar- and track-level controllability of the accompaniment with various textural attributes while keeping a homophonic texture.
With both subjective and objective evaluations, we show that our model outperforms style transfer models on a re-orchestration task in terms of generation quality and controllability.
Moreover, it can be adapted for a lead sheet orchestration task as a zero-shot learning model, achieving performance comparable to a model specifically trained for this task.
\end{abstract}

\section{Introduction}
\label{sec:intro}

Re-orchestration refers to the musical arrangement of an existing music piece for a different set of instruments~\cite{cacavas1975music}.
In the context of popular music, this notion is often associated with ``song covers''.
A key similarity between the original piece and its re-orchestration often lies in maintaining melodic fidelity. In Western music, which is predominantly homophonic, a primary melody is typically supported by an accompanying background~\cite{young2022form}.
Moreover, in the composition process, effective orchestration requires knowledge of writing for various instruments by combining their timbres, while being restricted by their physical limitations~\cite{adler1989study}.

Going further, re-orchestration extends beyond simply reassigning parts of the original piece to instruments in a new ensemble.  
It often involves altering the overall \emph{musical texture} of the piece to suit artistic goals or ensemble constraints.
Musical texture refers to how different musical streams are written, organized, and combined~\cite{huron89texture}.
An orchestral score can be described by global characteristics, such as instrument groupings or part diversity, and part-specific attributes such as rhythmicity or repetitiveness~\cite{le2022orchestral}.

In the field of symbolic music generation, re-orchestration can be considered as a \emph{style transfer} task, for which a model is designed to replicate a reference piece while altering high-level musical attributes.
However, existing style transfer systems
may be inadequate for specifically a re-orchestration task. 
They often focus on band arrangements~\cite{zhao2024structured,luo2024bandcontrolnet} which restricts the instrument choices to a fixed and small ensemble and does not allow fine-grained selection of instrumentation.
Moreover, these systems often overlook or even disregard the melodic fidelity of the generated content. For example, according to \figaro~\cite{von2022figaro}: ``some salient features such as melodies are often not preserved''.

Beyond the instrumentation choice, re-orchestration implies textural controls, for which style transfer systems have also been implemented.
This control is often performed at a piece-level~\cite{lu2023musecoco} or bar-level~\cite{wu2023musemorphose}.
For orchestral music -- more generally, multi-track music -- such control can also occur at the track level.

\begin{figure}[t]
    \centering
    \includegraphics[trim={.25cm .3cm .2cm .2cm},clip,width=\columnwidth]{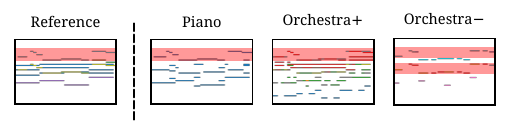}
    \caption{
        \meteor's re-orchestration task. 
        The model can re-orchestrate a reference for multiple instrumentations (\emph{e.g.} solo piano, or orchestra) 
        with texture controls, with more (orchestra+) or less (orchestra-) ``polyphonicity'' and ``rhythmic intensity'' (cf. Section~\ref{sec:textural_features}). 
        The models ensures
        melodic fidelity (red highlight) with fine-grained controls (melodic instrument choice and pitch range).
    }
    \label{fig:presentation_model}
\end{figure}

\renewcommand{\xmark}{{\color{gray}\ding{55}}}
\newcommand{\remark}[1]{{\it\color{black} #1}}
\newcommand{\asterisk}[1]{{\scriptsize #1$\ssymbol{2}$}}
\newcommand{\asteriskk}[1]{{\scriptsize #1$\ssymbol{7}$}}
\newcommand{\idptrk}{{\it\color{black}\cmark~{\scriptsize (ind. track)$\ssymbol{1}$}}}
\newcommand{\modelcite}[2]{#1~{\scriptsize #2}}

\begin{table*}
    \centering
    \begin{tabular}{lccccccc}
        \toprule
        \multirow{2}{*}{Model} & \multirow{2}{*}{\makecell{Multi-\\track}} & 
        \multicolumn{2}{c}{Texture controllability} & \multirow{2}{*}{\makecell{Melodic\\fidelity}} & \multicolumn{2}{c}{Instrument choice} &
        \multirow{2}{*}{\makecell{Open-\\source\raisebox{.25em}{\scriptsize 1}}} \\
        \cmidrule{3-4} \cmidrule{6-7} 
        & & Track-level & Bar-level & & Full ensemble & Melody  \\
        \midrule
        \modelcite{MuseMorphose}{\cite{wu2023musemorphose}} & 
            \xmark & 
            \xmark & 
            \cmark & 
            \xmark & 
            \xmark & 
            \xmark & 
            \cmark \\
        \modelcite{MuseBarControl}{\cite{shu2024musebarcontrol}} & 
            \xmark & 
            \xmark & 
            \cmark &
            \xmark &
            \xmark & 
            \xmark & 
            \xmark \\
        \modelcite{FIGARO}{\cite{von2022figaro}} &
            \cmark & 
            \cmark & 
            \cmark &
            \xmark & 
            \xmark & 
            \xmark & 
            \cmark \\
        \modelcite{PopMAG}{\cite{ren2020popmag}} & 
            \cmark & 
            \xmark & 
            \xmark & 
            \idptrk & 
            \remark{\asterisk{fixed (6)}} & 
            \xmark & 
            \xmark \\
        \modelcite{GetMUSIC}{\cite{lv2023getmusic}} & 
            \cmark & 
            \xmark & 
            \xmark &
            \idptrk &
            \remark{\asterisk{fixed (5)}} & 
            \xmark & 
            \xmark \\
        \modelcite{BandControlNet}{\cite{luo2024bandcontrolnet}} & 
            \cmark & 
            \cmark & 
            \cmark & 
            \idptrk & 
            \remark{\asterisk{fixed (6)}} & 
            \xmark & 
            \xmark \\
        \modelcite{AccoMontage-band}{\cite{zhao2024structured}} & 
            \cmark & 
            \xmark & 
            \remark{\asteriskk{implicit}} & 
            \idptrk & 
            \remark{\asteriskk{implicit}} & 
            \xmark & 
            \cmark \\
        \midrule
        \textbf{METEOR} (ours) & \cmark & \cmark & \cmark & \cmark & \cmark  & \cmark & \cmark \\
        \bottomrule
    \end{tabular}
    \caption{
        Models related to the style transfer sub-tasks performed by METEOR. 
        $(^{1})$ We consider models to be open-source when both the code and trained models are publicly available.
        $(\ssymbol{1})$ The melody is added \textit{a posteriori} as an independent track, in contrast with METEOR where the melodic instrument is chosen \emph{among} the chosen instrumentation.
        $(\ssymbol{7})$ This model mimics the texture and the instrumentation of an \textit{already existing} source: the choices are not explicit.
        $(\ssymbol{2})$ These models only handle a \textit{fixed} number of instrument types (\eg 5 or 6).
    }
    \label{tab:models_sota}
\end{table*}

In this study, we present \meteor, a model for \textbf{Me}lody-aware \textbf{Te}xture-controllable re-\textbf{Or}chestration (Section~\ref{sec:methods}). The model is designed to achieve the following (Figure~\ref{fig:presentation_model}): 
\begin{itemize}
    \item Multi-track music re-orchestration: the model automatically orchestrates a reference multi-track piece, with the instrumentation possibly specified by the user.
    \item Texture-controllability: textural attributes can be controlled at both bar and track levels.
    \item Melodic fidelity: the melody is preserved in the re-orchestrated piece, with the option for the user to select the melodic instrument.
\end{itemize}
To our best knowledge, %
\meteor is the first deep generative model that offers 
both \emph{instrumental} and \emph{texture-based} style transfers with melodic fidelity.
Moreover, we show that our model can perform a lead sheet orchestration task without further training in a zero-shot manner.
The main approach relies on an extension of MuseMorphose, a Transformer-based VAE, with bar- and track-level token constraints and inference guidance for melodic fidelity.
Section~\ref{sec:evaluation} provides an objective evaluation demonstrating \meteor's effectiveness in bar- and track-level controllability, melodic fidelity, and melodic instrument playability. 
A subjective evaluation further supports that it generates higher-quality re-orchestrations than baseline models.
We share audio extracts of generations on a demo page %
and open source code and model weights\footnote{\label{demopage}\url{https://github.com/dinhviettoanle/meteor}}.

\section{Related Works}
\label{sec:related_works}

Re-orchestration is a task which can be associated with multi-track music \textit{style transfer}, which aims at generating a multi-track piece by taking a multi-track reference and altering musical characteristics to reflect a specific style~\cite{dai2018music}.
Style transfer can refer to composer style transfer, where a music style is applied to a reference content~\cite{cifka2020groove2groove}. 
Though, our study focuses on two types of music style transfers: \emph{instrumental} style transfer, where the instrumentation of the reference piece is altered and \emph{texture-based} style transfer, where high-level musical features from the reference are adjusted to generate a new piece.
We specifically explore these tasks within the context of 
\textit{homophonic} music, which consists of a melody supported by an accompaniment.
Multiple models have been developed to address sub-tasks, with their strengths and limitations summarized in Table~\ref{tab:models_sota}.

\subsection{Multi-track Homophonic Music Generation}

Several models have been developed for multi-track music free generation~\cite{ens2020mmm,liu2022symphony,dong2023mmt} which can generate music without an initial musical reference. 
Comparatively, few studies have explicitly addressed the task of re-orchestration~\cite{von2022figaro}.
Closest style transfer models for this task focus on band arrangements~\cite{ren2020popmag,lv2023getmusic,luo2024bandcontrolnet,zhao2024structured}.
Such models usually only consider a fixed-number instrumental ensemble composed of generic instruments, such as drums, piano, or strings.
Moreover, while band music is usually written using a homophonic texture, defined as a primary melody supported by an accompaniment~\cite{benward2018music}, the melodic part is often overlooked.
These models either discard the melodic content~\cite{von2022figaro} or only generate the accompaniment and insert the melodic content \emph{a posteriori} into a track played by a \emph{fixed} instrument such as a synthesizer~\cite{luo2024bandcontrolnet} or a ``lead'' track~\cite{zhao2024structured}, without strict physical restrictions like its ambitus or register.
However, in styles such as Western classical orchestral music, the melody is assigned to a specific instrument or a group of instruments which can change throughout the piece to achieve particular timbre effects~\cite{adler1989study}. 
In such cases, the melody must respect the instrument's limitations, such as its range.

\begin{figure*}[t]
    \centering
    \includegraphics[width=0.76\linewidth]{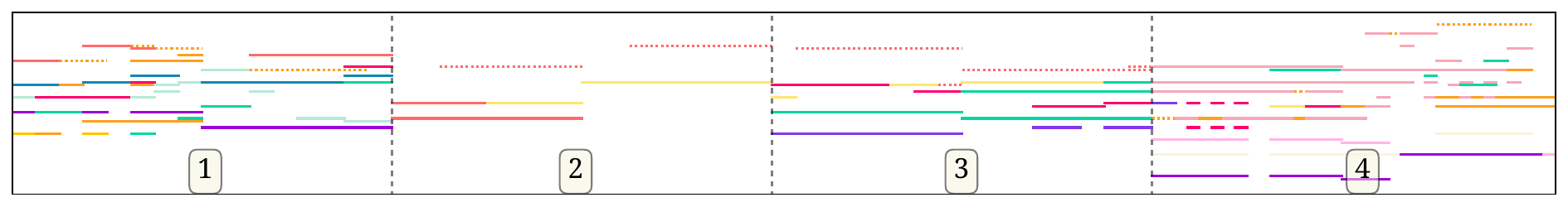}
    \raisebox{8pt}{\includegraphics[width=0.23\linewidth]{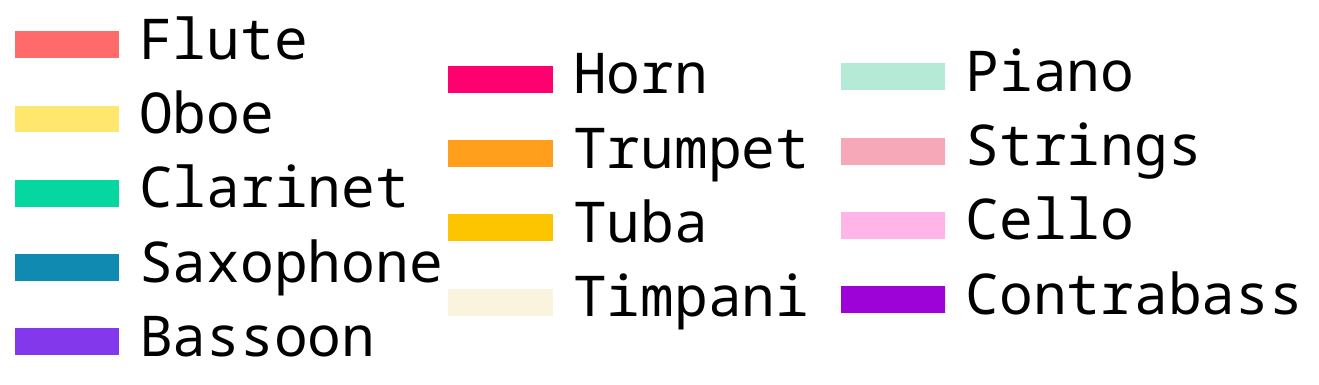}}
    \caption{
        Pianoroll of a 8-bar re-orchestration generation by \meteor with various textural and instrumentation constraints changing each 2 bars.
        Melody is dashed.
        (1) Automatic instrumentation, no textural changes.
        (2) Flute + oboe duet, with a melodic flute, low polyphonicity.
        (3) Wind quintet, with a melodic flute, low rhythmicity.
        (4) Classical orchestra, with a melodic trumpet, high rhythmicity and polyphonicity.
    }
    \label{fig:example_generation}
\end{figure*}

Adapting multi-track style transfer models for re-orchestration is not direct.
In particular, AccoMontage-band \cite{zhao2024structured}, designed for lead sheet band arrangement, suffers from several limitations for re-orchestration.
Beyond the \emph{a posteriori} melody insertion, it must rely on transcribing the multi-track input into a lead sheet used as input by the model. This dependency leads to challenges: the simplification of the textural content and a risk of transcription errors.

\subsection{Texture-Based Style Transfer}

Musical texture characterizes how musical streams are organized and describes their content~\cite{huron89texture}.
Texture-based style transfer systems often offer control on a reference piece over attributes such as rhythmic density~\cite{wu2023musemorphose} or pitch distributions~\cite{von2022figaro}.
Multiple levels of controllability can be defined \ie global, time-varying (often bar-level), and track-level (for multi-track music) controllabilities. Global features characterize the whole generated sequence; time-varying attributes only impact a single bar; and track-level controllability affects a single track either globally or locally at the bar level.

Bar-level textural controls are implemented by MuseMorphose~\cite{wu2023musemorphose} and MuseBarControl~\cite{shu2024musebarcontrol} for single-track piano music and \figaro~\cite{von2022figaro} for multi-track music. BandControlNet~\cite{luo2024bandcontrolnet} adds track-level controls. 
AccoMontage-band ~\cite{zhao2024structured} addresses a multi-track lead sheet arrangement task through texture transfer, but its controllability is limited. The model applies the texture of a ``texture donor'' to a musical content, restricting texture controllability to the set of \emph{pre-existing} texture donors.

MuseMorphose~\cite{wu2023musemorphose} appears to be a promising model for our task, offering fine-grained textural controls at a bar level. Though, a straightforward multi-track extension of the model may be insufficient for addressing the re-orchestration task, for example, due to the lack of melodic control. %
In contrast, the ideas introduced in Compose \& Embellish~\cite{wu2023compose}, a lead sheet piano arrangement model, provide insights that could address the issue of melodic control, in particularly through its approach of interleaving one-bar segments of melody and accompaniment, serving as an inspiration for the design of our model.

\section{Methods}
\label{sec:methods}

In this section, we introduce \meteor, 
a Transformer-based VAE model for multi-track re-orchestration 
with
instrumentation controllability,
bar- and track-wise texture controllability and melodic fidelity. %
We first present the musical attributes considered for textural controllability, and the technical contributions, particularly the tokenization strategies developed for this task. %

\subsection{Textural Attributes}
\label{sec:textural_features}

\meteor is a model designed for both instrumental and textural style transfer (Figure~\ref{fig:example_generation}). Specifically, its textural style transfer function enables the control of various textural attributes.
We consider two levels of controllability: ``bar-wise'' (\ie all tracks may be influenced by the control attribute) and ``bar- and track-wise'' (\ie each track can be individually controlled at a bar level).
We first consider bar-wise control attributes following \cite{wu2023musemorphose}.

\begin{itemize}
    \item \emph{Rhythmic intensity} (or \emph{rhythmicity}): number of sub-beats having at least one note played within a bar containing $B$ sub-beats, regardless of the track.
    With $\textbf{1}(\cdot)$ the indicator function, $
        s^\text{rhym} = \frac{1}{B} \sum_{b=1}^B \mathbf{1}(n_{\text{onset}, b} \geq 1)
    $.
    \item \emph{Polyphonicity}: average number of notes played (hit or held) during a sub-beat in a bar containing $B$ sub-beats, including all tracks.
    We consider $
        s^\text{poly} = \frac{1}{B} \sum_{b=1}^B (n_{\text{onset}, b} + n_{\text{hold}, b})
    $.
\end{itemize}

Each bar is characterized by a raw value of polyphonicity and rhythmicity.
These raw values are then split into 8 bins with a similar number of bars in each bin. These bins are set according to distribution of polyphonicity and rhythmicity values in the dataset.

For finer-grained control, we propose ``bar-wise and track-wise'' control attributes aiming at controlling each instrument individually among those initially selected at a bar level.

\begin{itemize}
    \item \emph{Average pitch}: average pitch of the set of pitches $\{p_1, \ldots, p_M\}$ played in a track $t$ in a bar, expressed in MIDI value and rounded to the nearest ten. 
    \begin{equation*}
        p^\text{avg}_\text{t} = \text{round}\left(\frac{1}{M} \sum_{i=1}^{M} p_i, 10\right)
    \end{equation*}
    Levels of average pitches are thus divided into 13 classes, spanning from 10 to 130.
    For instance, this attribute can be used to assign high register to melodic instruments, and low register for bass parts.
    \item \emph{Pitch diversity}: number of different pitch classes played in a track in a bar. 
    \begin{equation*}
        p^\text{diversity}_\text{t} = \left| \big\{ p_i \mod 12 \,\big|\, i = 1, 2, \dots, M \big\} \right|
    \end{equation*}
    Levels of pitch diversity are divided into 13 classes, spanning from 0 to 12.
    Low pitch diversity can relate to bass parts, repeated notes or arpeggios, while high pitch diversity can encourage passing notes, embellishments or extended chords.
\end{itemize}

\subsection{Tokenization, Model \& Control Strategies}
\label{sec:tokenization_models}

\begin{figure}
    \centering
    \fontfamily{lmss}\selectfont
    \resizebox{\linewidth}{!}{\input{figs/tikz_model_architecture_full_vertical_larger.tex}}
    \caption{
        Architecture of \meteor, based on MuseMorphose.
        The musical content in each bar is preceded by a \emph{header} describing the playing instruments in this bar and track-wise controls.
        During training, the model is trained to reconstruct $K$ bars. 
        At inference time, the user can specify different headers for each bar and starts the generation of $N$ bars starting from bar $i < K$ (\ie the user can ask to generate only from a sub-part of the full piece).
        The inference is guided with melody constraints at a beat level.
    }
    \label{fig:model_architecture}
\end{figure}

\meteor's architecture is based on MuseMorphose~\cite{wu2023musemorphose}, originally developed for piano style transfer.
The model implements a VAE based on Transformers encoders and decoders (Figure~\ref{fig:model_architecture}).
We first extend its initial REMI tokenization~\cite{huang2020pop} using the REMI+ tokenization~\cite{von2022figaro} which handles multi-track music.
Based on early experiments, we implement REMI+ using a vertical parsing\footnote{\url{https://musiclang.github.io/tokenizer}}, where notes are grouped and ordered based on time rather than track.
We also rely on a ``pitch class + octave encoding'' of the pitches~\cite{li2023pitchclass} instead of absolute MIDI values, in particular, to handle melodies independently of the original octave register.

For instrumentation controllabillity, the user can select the playing instruments from a subset of 64 instruments
defined in~\cite{dong2023mmt} or the ensemble can be automatically defined by the model.
Instrument selection is handled through \token{DescriptionTrack-[track]} tokens that indicate instruments playing in a bar which are added 
in a \emph{header} at the start of each bar in the token sequence.

For texture controllability, the model implements multiple controls over various textural attributes (Section~\ref{sec:textural_features}).
For bar- and track-wise controls, \token{PitchAvg-[track]-[level]} tokens and \token{PitchDiversity-[track]-[level]} tokens are added jointly in this header to describe the average pitch and pitch diversity level of each track.
A token sequence is shown in Figure~\ref{fig:tokenization}.
Following MuseMorphose, the bar-wise polyphonicity and rhythmicity classes are encoded in a separate sequence of bar-level conditions, which is embedded and concatenated with the latent vector and used as condition in the decoder through an ``in-attention'' mechanism.

Regarding the overall training process, the model is trained as an end-to-end model on the SymphonyNet dataset composed of 46k multi-track pieces~\cite{liu2022symphony}. 
Similar to MuseMorphose, the loss function used is a $\beta$-VAE objective with free bits. 
The resulting model is 67M parameter-large and is trained for one week on a single RTX 6000 24GB GPU.

\subsection{Inference Guidance for Melodic Fidelity}
\label{sec:melodic_fidelity}

Melodies are crucial elements in music, as they often make a piece %
easily recognizable~\cite{stefani1987melody}. 
Thus, a key focus of our model is melodic fidelity, ensuring that the original melody is preserved in the generated extract, with possibly different textures in the accompaniment parts. 
Models preserving the melody often insert \emph{a posteriori} a track containing the melody played by a generic instrument (\eg synthesizer),
which prevents any melodic ornamentation~\cite{le2022orchestral} and restricts its integration into the queried ensemble. 
Thus, we propose an \emph{inference guidance} process designed to ensure the melodic fidelity in a more flexible way in the generation.

First, the melody is identified in the original piece during a pre-processing step using a bar-wise and track-wise skyline algorithm. The melody in each bar is estimated as being the track with the highest average pitch within that bar\footnote{This assumption is a compromise as the melody can possibly be misdetected (\eg melodic bassoon or cello). Further improvements may be implemented with track role identification~\cite{guo2019midi}.}.

The instrument playing the melody is first chosen by the model or can be specified by the user. In particular, the model or the user may choose to use different instruments to play the melody in different bars of the generated piece.
The melody notes are then generated alongside with the re-orchestration using \emph{inference guidance}: tokens identified as melody in the original piece are treated as beat-level conditions at inference time. 
Following Figure~\ref{fig:model_architecture} (top left), after a \token{Bar} token and the enforced \emph{header} describing this bar, each \token{Sub-beat} token generated by the model is followed by an enforced \token{Track} token corresponding to the chosen melodic instrument, along with the tokens corresponding to the melody note played at this time position (\ie pitch class, octave, duration, and velocity). %
The next tokens (\ie all the accompaniment tokens until the next melodic tokens) are then generated auto-regressively. 
In particular, we do not restrict the model to generate additional notes played by the melodic instrument.
In further experiments (see Table~\ref{tab:octave}), we allow the model to infer \token{Octave} tokens to evaluate its relation with the instruments' register.

\begin{figure}
    \centering
    \fontfamily{lmss}\selectfont
    \resizebox{\linewidth}{!}{\input{figs/tikz_token_sequence_pitchclass_short.tex}}
    \caption{
        Example of token sequence for a bar with a violin and a flute. 
        As indicated in the \emph{bar header}, the violin plays in a medium register and has a high pitch diversity,
         while the flute is in the upper range, with a low pitch diversity.
    }
    \label{fig:tokenization}
\end{figure}

\subsection{Zero-Shot Lead Sheet Orchestration}

While \meteor has been specifically trained for a re-orchestration task, 
if can be adapted into a lead sheet orchestration model without requiring further training, effectively performing as a zero-shot learning model.
The model takes as input a lead sheet provided as a multi-track MIDI file, composed of a melodic track and a second track with block chords. 
By interpreting the lead sheet as a low-rhythmicity multi-track piece, \meteor is able to orchestrate this lead sheet with specific instruments by increasing the rhythmicity.

\section{Evaluation}
\label{sec:evaluation}

In this section, we first present an objective evaluation to assess our model's performance in terms of fidelity and controllability. 
This objective evaluation is then supported by a user study conducted as a subjective evaluation.

\subsection{Baseline Models}
\label{sec:baseline_models}

We compare \meteor with two open-source and state-of-the-art style transfer models %
and adapt them as multi-track re-orchestration models:
\begin{itemize}
    \item \figaro~\cite{von2022figaro}: This multi-track style transfer model can directly perform the re-orchestration task. For the evaluation, we focus on the proposed ``note density'' controls, which corresponds to the rhythmic intensity in our work.
    \item AccoMontage-band~\cite{zhao2024structured}: This model is originally designed to take a lead sheet as input and generate a multi-track pop band arrangement. We adapt this model to evaluate its performance as a re-orchestration system. To this end, we first pre-process a multi-track input its lead sheet representation \ie we extract the melody as the skyline stream and the chords using the Chorder package\footnote{\url{https://github.com/joshuachang2311/chorder}}. This extracted lead sheet is then used as the input for the model which generates the re-orchestration of the initial input.
\end{itemize}

We also consider a multi-track extension of MuseMorphose~\cite{wu2023musemorphose}, initially developed for piano textural style transfer, in which the original REMI tokenization is replaced with a REMI+ tokenization.

For the objective metrics, we compare these baselines with two versions of our model: ``\meteor without inference guidance'',  which includes bar- and track-level controllability but with ablated melody constraints (Section~\ref{sec:melodic_fidelity}), and ``\meteor'' which includes these melody constraints.

\subsection{Objective Metrics}

We first consider objective metrics to evaluate the full piece fidelity with respect to the reference piece, both overall and specifically for the melody.
We also consider a metric to evaluate the instrument realisticness in terms of pitch distribution.

\begin{itemize}

\item \emph{Overall fidelity --} 
Following~\cite{von2022figaro}, we consider the overall fidelity as the chroma similarity between the original piece and the generation,
defined as the average of bar-wise cosine similarities between bar-wise chroma vectors.

\item \emph{Melodic fidelity --}
For a piece, let $X_{b, \text{mel}}$ the token sequence representing the melody in bar $b$. For a track $t$ in the generation, let $X_{b,t}$ the token sequence of one track $t$ at this bar $b$. We consider the Levenstein edit distance between two sequences $d(\cdot, \cdot)$ and normalize it so that $\lvert d(X_1, X_2) \rvert \leq 1$ for $X_1$ and $X_2$ two sequences. 
We define the melodic fidelity of a track $t$ at a bar $b$ as $d(X_{b, \text{mel}}, X_{b,t})$.
By taking the minimum of these distances among the tracks, we aim at selecting the track which is playing the melody within a bar.
Therefore, the smaller the distance, the greater the melodic fidelity.
Namely, we define the melodic fidelity $\varphi_{_b}$ 
at a bar $b$ as:
\begin{equation*}
    \varphi_{_b} = 1 - \min_{t \in \text{tracks}} d(X_{b, \text{mel}}, X_{b,t})
\end{equation*}
Finally, we define the melodic fidelity $\varphi_\text{mel} \in [0,1]$ of a full multi-track generation of $N$ bars as the average of these bar-wise fidelities:
    $\varphi_\text{mel} = \frac{1}{N}\sum_{b=1}^N \varphi_{_b}$

\item \emph{Pitch distribution similarity per instrument --}
To evaluate the re-orchestration instrumental realisticness, we compare the distribution of pitches per instruments between a generated content and a reference dataset.
Let $P_i$ (resp. $Q_i$) the distribution of pitches played by the instrument $i$ in a reference dataset\footnote{This reference dataset includes the SymphonyNet dataset and an equal number of pieces from the LakhMIDI dataset.} (resp. in the generated music). 
We consider that $i$ is among the $I$ available instruments. For $\text{JSD}(\cdot||\cdot)$ the Jensen–Shannon divergence, we define the instrument pitch distribution similarity $\rho \in [0, 1]$:
\begin{equation*}
    \rho
    = \frac{1}{I} \sum_{i=1}^T 
    \left(1 - \text{JSD}(D_i||Q_i)\right)
\end{equation*}

\end{itemize}

Regarding textural controllability, we then consider bar-level metrics for polyphonicity and rhythmicity and bar- and track-level metrics for average pitch and pitch diversity. 

\begin{itemize}
\item 
\emph{Bar-controllability --}
Polyphonicity and rhythmicity are evaluated at the bar level by including all tracks.
Following~\cite{wu2023musemorphose}, we consider the Spearman correlation between the user-specified polyphonicity or rhythmicity class and the class computed from the model generations given the user inputs.

\item 
\emph{Track-controllability --}
Average pitch and pitch diversity are also evaluated with a Spearman correlation between the user input and the class computed from the generation, for each track and each bar.

\end{itemize}

For this evaluation, each model generates 20 samples of 8 bars each, with the reference pieces and the control signals randomly selected and the instruments chosen by the models.

\newcommand{\rownum}[1]{{\tiny #1} \hspace{-8pt}}
\newcommand{\uarrow}{{\scriptsize $\uparrow$}}
\newcommand{\meanstd}[2]{{#1}~{\tiny $\pm$#2}}

\begin{table*}
    \centering
    \small
    \begin{tabular}{rlr|ccc|cc|cc}
         \toprule
        &   \multicolumn{2}{l|}{\multirow{2}{*}{Model}} & 
            \multirow{2}{*}{\makecell{Overall\\fidelity \uarrow}} & 
            \multirow{2}{*}{\makecell{Melodic\\fidelity \uarrow}} & 
            \multirow{2}{*}{\makecell{Instr. pitch\\ similarity \uarrow}} & 
            \multicolumn{2}{c|}{Bar-controllability \uarrow} & 
            \multicolumn{2}{c}{Track-controllability \uarrow} \\
        & & & & & & Rhyth. & Polyph. & Pitch diver. & Avg. pitch \\ 
         \midrule
        \rownum{1} & 
            \multicolumn{2}{l|}{\figaro} & 
            \meanstd{.735}{.24} & 
            \meanstd{.271}{.08} & 
            \meanstd{.617}{.14} & 
            .867  & 
            -- &  
            -- & 
            -- \\
        \rownum{2} & 
            \multicolumn{2}{l|}{AccoMontage-band} & 
            \meanstd{.756}{.10} & 
            \meanstd{.338$\ssymbol{1}$}{.09} & 
            \meanstd{.583}{.17} & 
            -- & 
            -- & 
            -- & 
            -- \\
        \midrule
        \rownum{3} & 
            \multicolumn{2}{l|}{Multi-track MuseMorphose} & 
            \meanstd{\textbf{.932}}{.10} & 
            \meanstd{.527}{.13}  & 
            \meanstd{.696}{.18} & 
            .941 & 
            .936 & 
            -- & 
            -- \\
        \rownum{4} & 
            \multicolumn{2}{l|}{\meteor~{\scriptsize (w/o inference guidance)}} & 
            \meanstd{.918}{.11} & 
            \meanstd{.491}{.16} & 
            \meanstd{.755}{.13} & 
            \textbf{.972} & 
            \textbf{.951} & 
            \textbf{.929} & 
            \textbf{.926} \\
        \rownum{5} & 
            \multicolumn{2}{l|}{\meteor} & 
            \meanstd{.927}{.10} & 
            \meanstd{\textbf{.632}}{.18} & 
            \meanstd{\textbf{.780}}{.12} & 
            .950 & 
            .932 & 
            .897  & 
            .821 \\
         \bottomrule
         \toprule
        \rownum{6} & \multirow{3}{*}{\makecell[l]{Multi-track\\MuseMorphose}} & \instrrst{\quad \textit{Flute-oboe duet}} & \instrrst{.888} & \instrrst{.479} & \instrrst{--} & \instrrst{.919} & \instrrst{.710} & \instrrst{--} & \instrrst{--} \\
        \rownum{7} & & \instrrst{\quad \textit{Woodwind quintet}} & \instrrst{.932} & \instrrst{.519} & \instrrst{--} & \instrrst{.956} & \instrrst{.875} & \instrrst{--} & \instrrst{--} \\
        \rownum{8} & & \instrrst{\quad \textit{Classical orchestra$\ssymbol{2}$}} & \instrrst{.947} & \instrrst{.511} & \instrrst{--} & \instrrst{.921} & \instrrst{.676} & \instrrst{--} & \instrrst{--} \\
         \midrule
        \rownum{9} & \multirow{3}{*}{\makecell[l]{\meteor \\ {\scriptsize (w/o infer. guidance)}}} & \instrrst{\quad \textit{Flute-oboe duet}} & \instrrst{.837} & \instrrst{.457} & \instrrst{--} & \instrrst{.967} & \instrrst{.782} & \instrrst{.949} & \instrrst{.873} \\
        \rownum{10} & & \instrrst{\quad \textit{Woodwind quintet}} & \instrrst{.903} & \instrrst{.519} & \instrrst{--} & \instrrst{.971} & \instrrst{.860} & \instrrst{.961} & \instrrst{.853} \\
        \rownum{11} & & \instrrst{\quad \textit{Classical orchestra$\ssymbol{2}$}} & \instrrst{.917} & \instrrst{.493} & \instrrst{--} & \instrrst{.975} & \instrrst{.898} & \instrrst{.958} & \instrrst{.798}\\
         \midrule
        \rownum{12} & \multirow{3}{*}{\meteor} & \instrrst{\quad \textit{Flute-oboe duet}} & \instrrst{.837} & \instrrst{.650} & \instrrst{--} & \instrrst{.936} & \instrrst{.715} & \instrrst{.786} & \instrrst{.711} \\
        \rownum{13} & & \instrrst{\quad \textit{Woodwind quintet}} & \instrrst{.909} & \instrrst{.651} & \instrrst{--} & \instrrst{.927} & \instrrst{.862} & \instrrst{.889} & \instrrst{.875}\\
        \rownum{14} & & \instrrst{\quad \textit{Classical orchestra$\ssymbol{2}$}} & \instrrst{.912} & \instrrst{.720} & \instrrst{--} & \instrrst{.953} & \instrrst{.867} & \instrrst{.909} & \instrrst{.780}\\
         \bottomrule
    \end{tabular}
    \caption{
        Objective metrics for the re-orchestration task, with automatic choice and user-defined ensembles. 
        $(\ssymbol{1})$ we evaluate the generated content only, without the inserted melodic track.
        $(\ssymbol{2})$ Classical orchestra includes 11 instruments (4 woodwinds, 2 brasses, timpani, 4 strings).
    }
    \label{tab:results}
\end{table*}

\subsection{Subjective Metrics}

Following these objective metrics, we conduct a user study to compare \meteor with the two baseline models. 
We evaluate the quality of the generations on the task of re-orchestration (multi-track to multi-track) and lead sheet orchestration (lead sheet to multi-track).
For both tasks, participants listen to a 8-bar long reference (multi-track piece or lead sheet) and samples generated by the 3 models (Section~\ref{sec:baseline_models}).
For the first task, they are asked to rate the generation contents on a 
6-point Likert 
scale from 0 (very low) to 5 (very high) based on the following criteria and guidelines: 
\begin{itemize}
    \item \textbf{Overall musicality}: how enjoyable is the music?
    \item \textbf{Naturalness of the generation}: to what degree does the piece meet your expectations for musical plausibility?
    \item \textbf{Textural fidelity with the reference}: how does the extract reflect the reference ``mood'' (calmness, energy...)?
    \item \textbf{Convincing use of instruments}: how well do the instruments blend together within the overall arrangement?
    \item \textbf{Content coherency with the reference}: how much do you recognize the reference by listening to the sample?
\end{itemize}
The same aspects are evaluated for the lead sheet orchestration task, without ``textural fidelity'' and with a ``creativity'' criterion (how inventive while being still pleasant to hear, is the audio extract). 
We let the model choose the melodic track automatically (or randomly for \figaro and AccoMontage-band) in the re-orchestration task.
Instead, for lead sheet orchestration, we insert \textit{a posteriori} the melodic track played by a synthesizer, following the method of AccoMontage-band. 
This ensures a fair comparison of all models in terms of melody perception by the listener, allowing for a focused comparison between the generated accompaniments.

The survey consists of 6 pieces for the re-orchestration task and 4 for lead sheet orchestration, chosen to ensure diversity. For each piece, the instrumentation is fixed for all models, including different cases: where the number of target instruments is smaller or greater than the source instruments. 
Each model generates four re-orchestrations for each 6 pieces. Participants are randomly assigned to one of the four groups, with each group evaluating a different set of samples.
A total of 24 participants for the re-orchestration task, and 13 for lead sheet orchestration have answered the survey. 
They have various musical backgrounds, from individuals with no musical experience (15\%) to professional musicians (8\%), with a majority of amateur (46\%) to intermediate musicians (31\%).

\subsection{Results}
\label{sec:results}

\paragraph{Objective evaluation}

Quantitative metrics 
are summarized in Table~\ref{tab:results} (rows 1--5).
MuseMorphose and the two versions of \meteor manage to outperform baseline models in all metrics. 
With \figaro, they outperform AccoMontage-band in pitch distribution fidelity, as both are trained on orchestral instruments while AccoMontage-band is trained on band instruments.
MuseMorphose and the two \meteors achieve comparable overall fidelities and adding melodic constraints naturally leads to an improvement in melodic fidelity. 
Though, the latter does not reach a perfect score, as \emph{inference guidance} does not prevent the melodic instrument from adding extra notes beyond the exact melody, a phenomenon which can be found in orchestral music, often referred to as ``decorative melody''~\cite{le2022orchestral}.
This increase in melodic fidelity results in a drop in controllability metrics compared to \meteor w/o melody. 
This may result from using independent control methods, either latent or token-based, for beat-, bar- and track-level attributes. 
The compatibility between latent space-based or token-based controls remains unexplored and could be investigated in future research to improve the understanding of controllable models.

\paragraph{Instrumentation impact} 
We further study the impact of the chosen instrumentation on our models' performances (Table~\ref{tab:results}, rows 6--14). 
We select three musical ensembles: woodwind duet, quintet,
and classical orchestra, assigning the melody to the flute in each case.
For \meteors, 
increasing the number of instruments helps the model maintaining better fidelity to the reference piece and improves bar-wise attributes.
With more instruments, the model has a larger instrumental flexibility and a broader range of options to assign each track a part that aligns with the control signals.
Moreover, all the models demonstrate better bar-wise polyphonicity controllability when the instrumentation is chosen automatically (rows 3--5) compared to each user-defined ensembles. In other words, they manage to effectively select the most suitable ensemble to match the requested polyphonicity.

\paragraph{Melodic instrument range playability}
We then study the playability of generations in terms of physical constraints of the melodic instrument. Unlike generic instruments such as synthesizers~\cite{luo2024bandcontrolnet,zhao2024structured}, orchestral instruments are limited in their range and usually play in a specific register~\cite{rimsky1964principles}.

To evaluate such range playability, we generate five extracts from the same original reference without textural control attributes and assign the melody to an instrument.
Based on our pitch class-based tokenization, we let the model infer the \token{Octave} tokens of the melody notes, while the other components (pitch class and duration) are enforced.
As presented in Table~\ref{tab:octave}, the model manages to generate instrumental parts which match with their usual register, with still a limited amount of out of range notes.
However, while the difference between the generated average pitch and its middle-range note is below a fourth for woodwinds and trumpet, the average generated pitch for the cello and the violin are much lower (\eg a sixth lower than the midpoint note of the violin's full register). Violin parts and, more generally, string parts, are indeed typically written below the extreme high register of the instrument~\cite[p.~52]{adler1989study}.

\newcommand{\register}[3]{#1 {\scriptsize (#2-#3)}}

\begin{table}
    \centering
    \scriptsize
    \begin{tabular}{lccc}
        \toprule
         \makecell[l]{Melodic\\instrument} & \makecell{Average note in instr. range\\(register bounds)} & \makecell{Average pitch\\in generations} & \makecell{Out of range\\generated notes}\\
         \midrule
         Flute & \register{F5}{B3}{C7} & {D5} & {0.0\%}\\
         Bassoon & \register{E3}{B$\flat$1}{B4} & {B$\flat$2} & {4.8\%} \\
         Trumpet & \register{A4}{F\#3}{C6} & {G4} & {4.3\%}\\
         Violin & \register{A5}{G3}{B7} & {C5} & {1.6\%} \\
         Cello & \register{B$\flat$3}{C2}{A5} & {F3} & {3.8\%}\\
         \bottomrule
        \end{tabular}
    \caption{
    Average pitch of melodic instruments with octave inference in the generated music compared to their real instrumental range.
    }
    \label{tab:octave}
\end{table}

\begin{figure}
    \centering
    \begin{subfigure}{\linewidth}
        \centering
        \includegraphics[width=\linewidth]{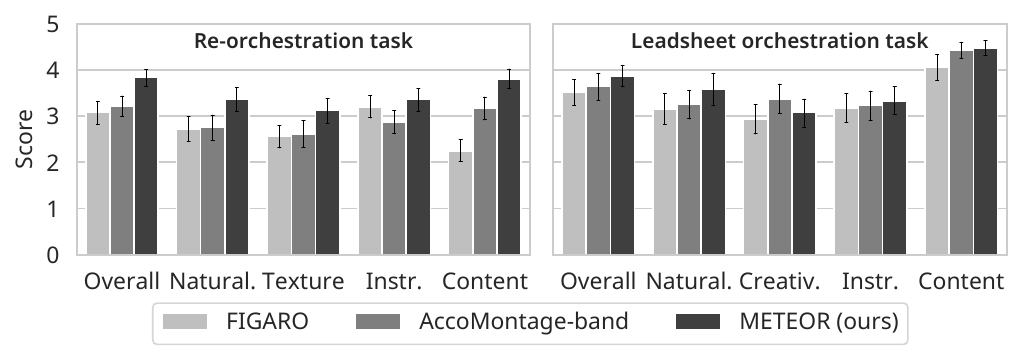}
        \caption{Average scores obtained on the re-orchestration and lead sheet orchestration tasks.}
        \label{figsub:user_study_results}
    \end{subfigure}
    
    \begin{subfigure}{\linewidth}
        \centering
        \includegraphics[width=.85\linewidth]{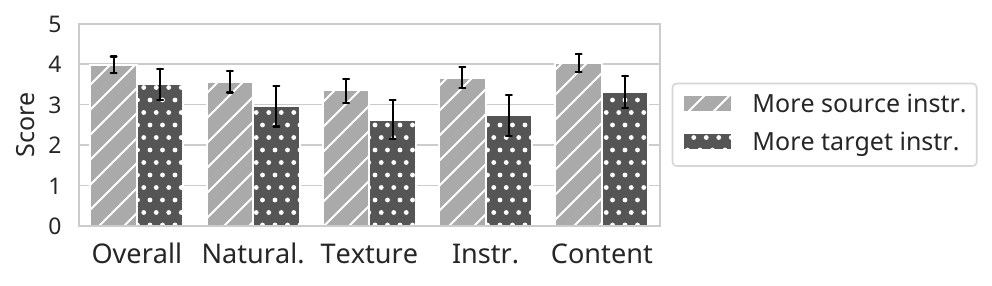}
        \caption{Impact of the number of source and target instruments on \meteor's re-orchestrations.}
        \label{figsub:user_study_instruments}
    \end{subfigure}
    \caption{
        Subjective evaluation results. A 6-point Likert scale ranging from 0 (very low preference) to 5 (very high preference) is used.
    }
    \label{fig:user_study}
\end{figure}

\paragraph{Subjective evaluation}
The results from our user study on each task and criterion are presented in Figure~\ref{fig:user_study}.

\textit{Re-orchestration task --}
\meteor outperforms in four of the five criteria on average (Figure~\ref{figsub:user_study_results}, left). In particular, it holds significant advantage over the two other models on the overall musicality and naturalness (t-test: $p < .01$ for both). Further analysis highlights notable insights on other criteria.
\begin{itemize}
\item Texture fidelity. \meteor achieves significantly better results than the baseline models, in particular, compared to AccoMontage-band ($p < .01$). 
This may be attributed to the lead sheet input which simplifies the original piece by reducing it to melody and chords, losing crucial textural characteristics and making it challenging for the model to generate a similar musical texture.
\item Instrumental use. \meteor and \figaro show comparable performances 
and both outperform AccoMontage-band. 
Given that the ensembles have been set to standard orchestral instruments, this shows that AccoMontage-band, which was trained with pop band instruments, can weakly adapt to unseen instruments.
However, when comparing scenarios with varying numbers of target instruments relative to source instruments (Figure~\ref{figsub:user_study_instruments}), \meteor performs better on instrumentation reduction and weaker when the target ensemble is larger than the reference on all criteria.
This may be attributed to the need for generating longer sequences for these larger ensembles, highlighting a potential limitation in the model's ability to capture long-term dependencies.
\item Content coherency. \figaro has an average score significantly lower than \meteor %
and AccoMontage-band %
($p<1\mathrm{e}{-6}$).
As noted in the original study, \figaro often fails to preserve the melody, highlighting that content coherency is strongly influenced by the retention of the melodic line. 
This effect is supported by the observation that its content fidelity is more comparable to other models in the lead sheet orchestration task, where the melody is inserted unchanged \textit{a posteriori}.

\end{itemize}

\textit{Lead sheet orchestration task --}
Across all metrics, \meteor achieves performance ranging from comparable to better than the other models (Figure~\ref{fig:user_study}, right).
In particular, while AccoMontage-band has been specifically trained on this task, it only outperforms \meteor on average on the creativity criterion. 
This zero-shot learning ability highlights \meteor's versatility in performing tasks closely related to orchestration with comparable performances with state-of-the-art models.

\section{Conclusion \& Future Directions}
\label{sec:conclusion}

In this study, we present \meteor, a model for texture-controllable multi-track style transfer with a focus on melodic fidelity specifically trained for a task of re-orchestration. 
The model performs this task through token constraints at a bar- and track-level, with inference guidance for melodic fidelity.
On a re-orchestration task, \meteor outperforms multi-track style transfer models on subjective and objective evaluations.
We show that our model can be adapted into a lead sheet orchestrator and is comparable to a model trained for this task.

\paragraph{Limitations \& future directions}
Our study focuses on bar- and track-level controllability, excluding piece-level controllability~\cite{lu2023musecoco}, 
which consequently disregards high-level structures, such as repeated musical phrases, which are fundamental in music composition~\cite{shih2022theme}. 
Future work towards multi-level multi-track style transfer may include a model able to perform style transfer at these three levels.
Such a controllable model could be integrated into an orchestrator's workflow as a co-creative tool, allowing both broad orchestration drafts and detailed refinements.
From a musical perspective, although \meteor succeeds in ensuring that melodic instruments fit their range constraints, their technical playability (\eg convenient fingerings, breath considerations, logical articulations) have not been thoroughly studied and are systematically overlooked in music generation studies. 
Ensuring playability in relation to instrumental constraints, timbre effects, and instrument groupings~\cite{goodchild2018perceptual} would be a significant advancement towards automatic humanly playable orchestration.

\section*{Ethical Statement}

Our work focuses on automatic music generation, raising potential concerns about the ownership of the generated content. 
Though, our study emphasizes human-machine co-creativity, particularly by enabling fine-grained control over the textural and instrumental properties of the generated content.
These controls are still limited regarding the style of music due to the choice of the training dataset which inherently exhibits a bias towards a Western instrumentation and a Western tonal style of music.%

Moreover, our study's evaluation relies on a survey presented as a user listening test. In this survey, no personal information was retrieved and the data was not used for other purposes than the current study.

Finally, our study is based on a deep learning approach, which may have an energy consumption impact due to the computational power required for model development, training, and evaluation. 
Although we did not precisely monitor any hardware power consumption during this study, an approximation\footnote{\url{https://mlco2.github.io/impact}}~\cite{lacoste2019quantifying} of a training of our model, limited to a duration of one week on our hardware, reaches a consumption of 7~$\text{kgCO}_{2}$eq.

\section*{Acknowledgements}
The work of Dinh-Viet-Toan Le was supported by grant ANR-20-THIA-0014 program ``AI\_PhD@Lille'' 
and a bourse MERMOZ from Région Hauts-de-France. 
The work of Yi-Hsuan Yang was supported by a grant from the National Science and Technology Council of Taiwan (NSTC 112-2222-E002-005-MY2) and a grant from Google Asia Pacific.
The authors would like to thank 
the anonymous reviewers for their valuable feedback which helped improve the article,
Ken Déguernel and Chih-Pin Tan for their comments on earlier versions of this manuscript, 
and Louis Bigo and Mikaela Keller for fruitful discussions.

\bibliographystyle{named}
\bibliography{refs}

\end{document}